\documentclass[12pt]{iopart}

\usepackage{iopams}
\usepackage{graphicx}

\newcommand{\Fi}[1]   {Fig.~\ref{#1}}

\newcommand{\agev}    {\mbox{$A$~GeV}}               

\newcommand{\rb}[1]   {\mbox{\textrm{\scriptsize #1}}}
\newcommand{\rbt}[1]  {\mbox{\textrm{\tiny #1}}}
\newcommand{\sqrts}   {\ensuremath{\sqrt{s_{_{\rbt{NN}}}}}}

\newcommand{\lam}     {\ensuremath{\Lambda}}
\newcommand{\lab}     {\ensuremath{\bar{\Lambda}}}  
\newcommand{\xim}     {\ensuremath{\Xi^{-}}}
\newcommand{\xizero}  {\ensuremath{\Xi^{0}}}
\newcommand{\xip}     {\ensuremath{\bar{\Xi}^{+}}}

\newcommand{\nwound}  {\ensuremath{\langle N_{\rb{w}} \rangle}}

\newcommand{\yproj}   {\ensuremath{y_{\rb{p}}}}
\newcommand{\deltay}  {\ensuremath{\langle \delta y \rangle}}
\newcommand{\rdely}   {\ensuremath{\langle \delta y \rangle / y_{\rb{p}}}}
\newcommand{\dnetbar} {\ensuremath{\textrm{d}N_{(\rbt{B}-\bar{\rbt{B}})}/\textrm{d}y}}

\begin{document}

\title[]{Centrality and Energy Dependence of Proton, Light Fragment and
Hyperon Production}

\author{Christoph Blume for the NA49 Collaboration}

\address{Institut f\"ur Kernphysik der J.W.~Goethe Universit\"at, \\
60438 Frankfurt am Main, Germany}

\begin{abstract}
Recent results of the NA49 collaboration are discussed.  These include
the energy dependence of stopping and the production of the light 
fragments $t$ and $^{3}$He.  New data on the system size dependence
of hyperon production at 40$A$ and 158\agev\ are also presented.
\end{abstract}




\section{Introduction}

In the recent years the NA49 experiment has collected data on Pb+Pb 
collisions at beam energies between 20$A$ and 158\agev\ with the objective 
to cover the critical region of energy densities where the expected 
phase transition from a deconfined phase might occur in the early stage 
of the reactions. 
NA49 is a fixed target experiment at the CERN SPS.
Details on the experimental setup can be found in \cite{NA49NIM}.

\section{Energy Dependence of Stopping}

\begin{figure}[htb]
\begin{center}
\begin{minipage}[b]{55mm}
\begin{center}
\includegraphics[height=65mm]{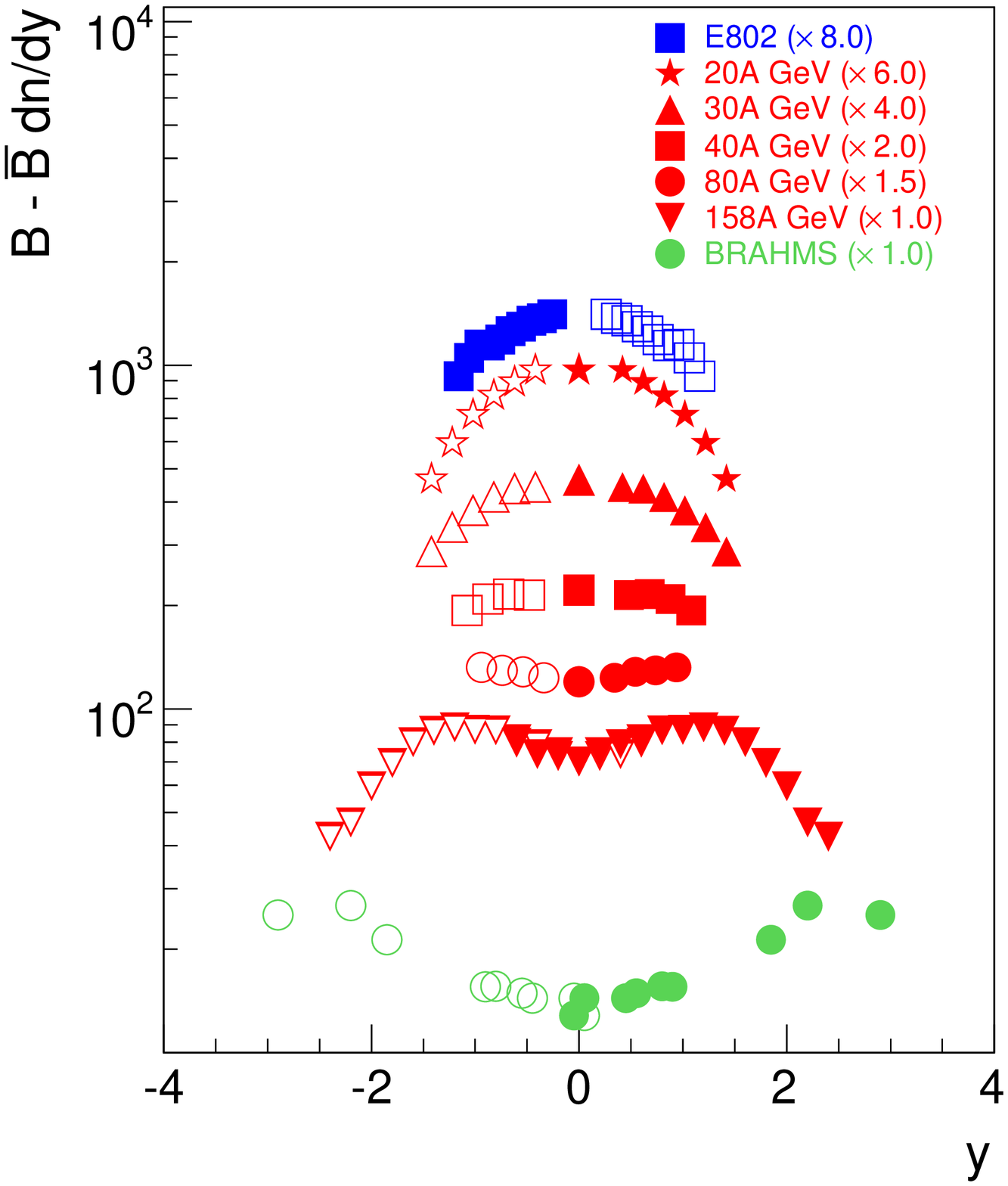}
\end{center}
\end{minipage}
\begin{minipage}[b]{85mm}
\begin{center}
\includegraphics[height=65mm]{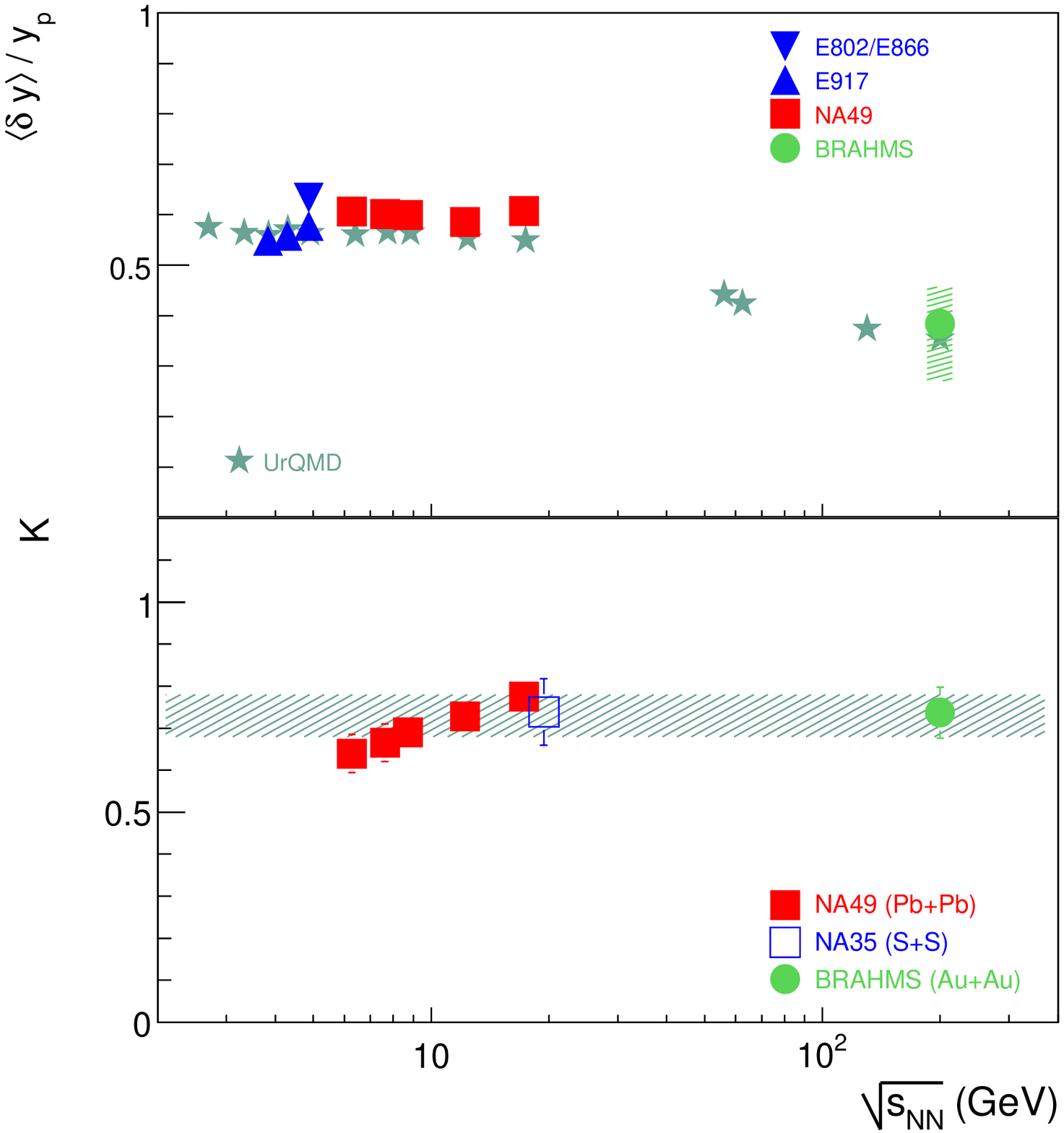}
\end{center}
\end{minipage}
\end{center}
\caption{Left panel: The rapidity distributions of net-baryons at SPS
energies~\cite{NA49STOP} together with results from the 
AGS~\cite{E802PROT} and from RHIC~\cite{BRMSSTOP} for central 
Pb+Pb(Au+Au) collisions.
Right panel: The relative rapidity shift \rdely\ as a function of the
projectile rapidity \yproj\ \cite{NA49STOP,BRMSSTOP,E917STOP,E802STOP}
(upper part).  Also shown are results for the UrQMD model \cite{URQMD}.
The lower part summarizes the \sqrts-dependence of the inelasticity
$K$, including NA35 data for central S+S reactions~\cite{NA35}.}
\label{fig:stopping}
\end{figure}

New results on rapidity spectra of (anti-)protons in central Pb+Pb
reactions at 20$A$ - 80\agev\ in combination with previously published 
results~\cite{NA49PROT,NA49STOP} allow to study the energy evolution 
of stopping.  Based on the measured rapidity spectra for p, 
$\bar{\textrm{p}}$, \lam, \lab, \xim, and \xip, all corrected for feed down
from weak decays, the net-baryon distributions \dnetbar\ are constructed.  
The contribution of unmeasured baryons (n, $\Sigma^{\pm}$, \xizero) is 
estimated using the results of a statistical hadron gas model \cite{BECATT1}.  
In the SPS energy region a clear evolution of the shape can be observed 
(see \Fi{fig:stopping} left).  From these distributions an averaged rapidity 
shift \deltay\ can be derived:
\begin{equation}
\langle \delta y \rangle = y_{\rb{p}} 
                         - \frac{2}{N_{\rb{part}}} \int_{0}^{y_{p}}
                           y \: 
                           \frac{\textrm{d}N_{(\rbt{B}-\bar{\rbt{B}})}}
                                {\textrm{d}y} \: 
                           \textrm{d}y,
\end{equation}
where \yproj\ is the projectile rapidity and $N_{\rb{part}}$ the number
of participating nucleons.  At AGS and SPS energies a value of 
$\rdely \approx 0.6$ is observed, which drops to $\rdely \approx 0.4$ 
at \sqrts~=~200~GeV (see \Fi{fig:stopping} right).
The measurements agree quite well with the UrQMD model \cite{URQMD}.
Using \dnetbar\ and the measured $\langle m_{\rb{t}} \rangle$ the 
inelastic energy per net-baryon:
\begin{equation}
E_{\rb{inel}} = \frac{\sqrt{s_{\rbt{NN}}}}{2} - 
                \frac{1}{N_{(\rbt{B}-\bar{\rbt{B}})}}
                \int_{-y_{p}}^{y_{p}} \langle m_{\rb{t}} \rangle \:
                \frac{\textrm{d}N_{(\rbt{B}-\bar{\rbt{B}})}}{\textrm{d}y} 
                \: \cosh y \: \textrm{d}y
\end{equation}
and the inelasticity $K = 2 \:E_{\rb{inel}} / (\sqrt{s_{\rbt{NN}}} - 
2 m_{\rb{p}})$ can be calculated.  $K$ is approximatly energy independent 
with a value of $K \approx 0.7 - 0.8$ (see right panel of \Fi{fig:stopping}).

\section{Light Fragments}

NA49 has measured the energy dependence of tritons at midrapidity (not 
discussed here) and $^{3}$He in a wide rapidity range.  The left panel 
of \Fi{fig:fragments} shows the rapidity distributions for $^{3}$He at 
four different beam energies.  The spectra have a similar concave shape 
at all investigated energies.  This is in remarkable constrast to the 
case of protons, where the shape of the rapidity distributions is strongly 
energy dependent.  By extrapolating the measured distributions total 
multiplicities can be determined (see right panel of \Fi{fig:fragments}).  
The values agree very well with a prediction of a statistical hadron 
gas model \cite{BECATT1}, similar to what was suggested for AGS
energies in \cite{PBM}.  

\begin{figure}[htb]
\begin{center}
\begin{minipage}[b]{70mm}
\begin{center}
\includegraphics[height=60mm]{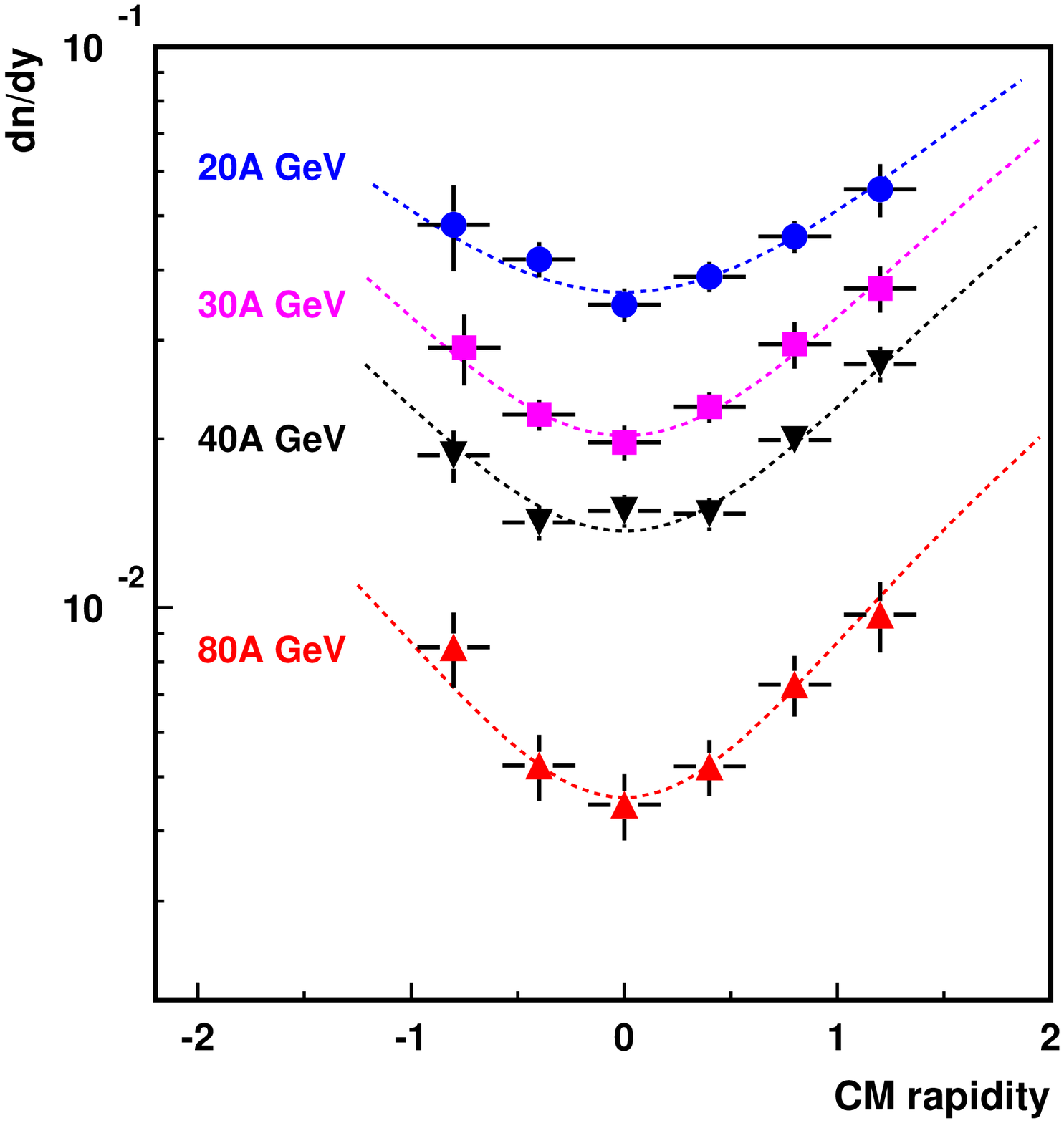}
\end{center}
\end{minipage}
\begin{minipage}[b]{70mm}
\begin{center}
\includegraphics[height=60mm]{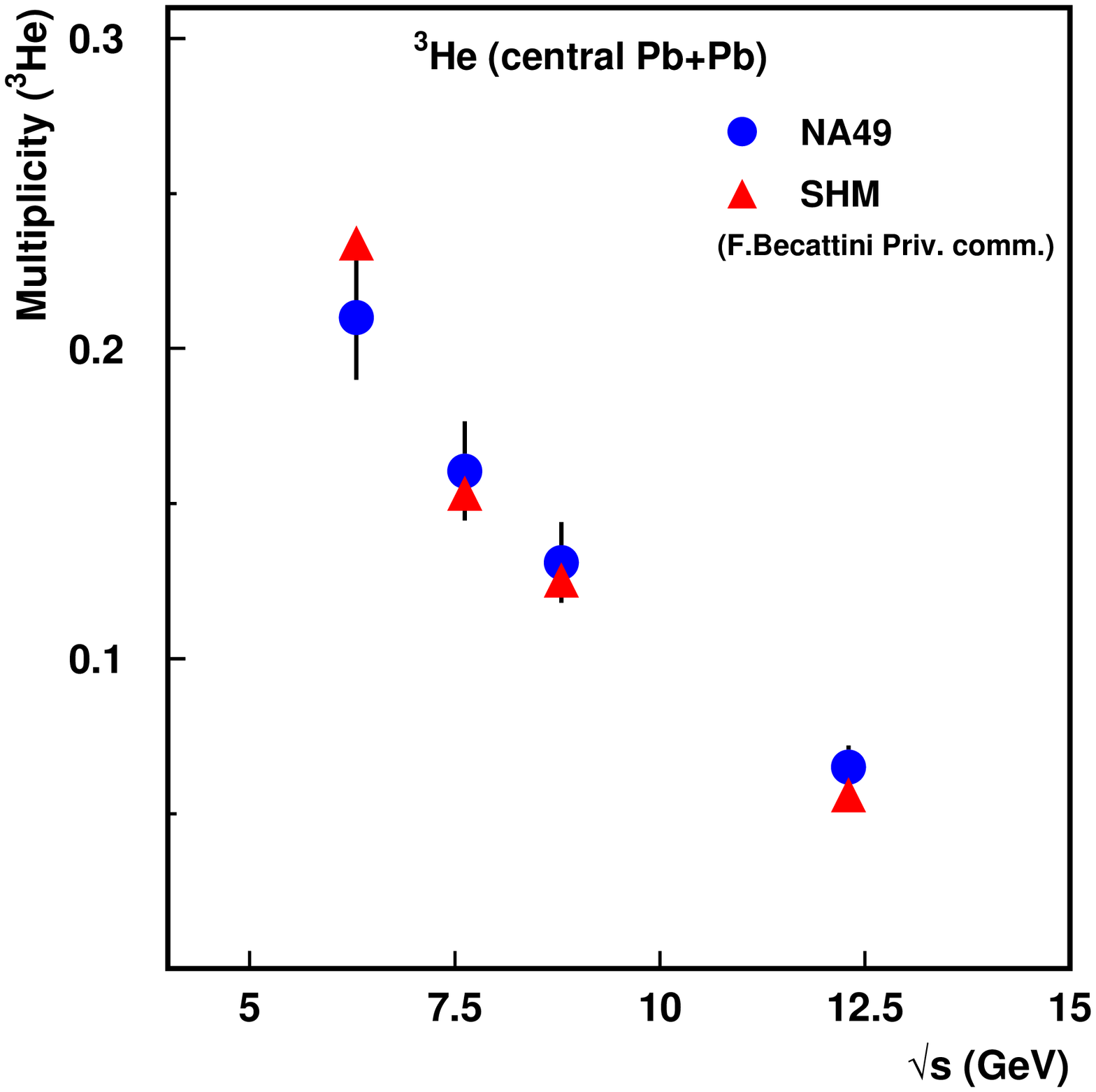}
\end{center}
\end{minipage}
\end{center}
\caption{Left panel: Rapdity distributions of $^{3}$He in central 
Pb+Pb collisions at 20$A$, 30$A$, 40$A$, and 80\agev.  The dashed
lines are fits with a parabola which is used to extract the total
multiplicities.
Right panel: The total multiplicity of $^{3}$He in central Pb+Pb
collisions as a function of \sqrts\ (solid points).  The triangles
represent predictions of a statistical hadron gas model
\cite{BECATT1}.}
\label{fig:fragments}
\end{figure}

\section{System Size Dependence of Hyperons}

\begin{figure}[htb]
\begin{minipage}[b]{50mm}
\begin{center}
\includegraphics[height=60mm]{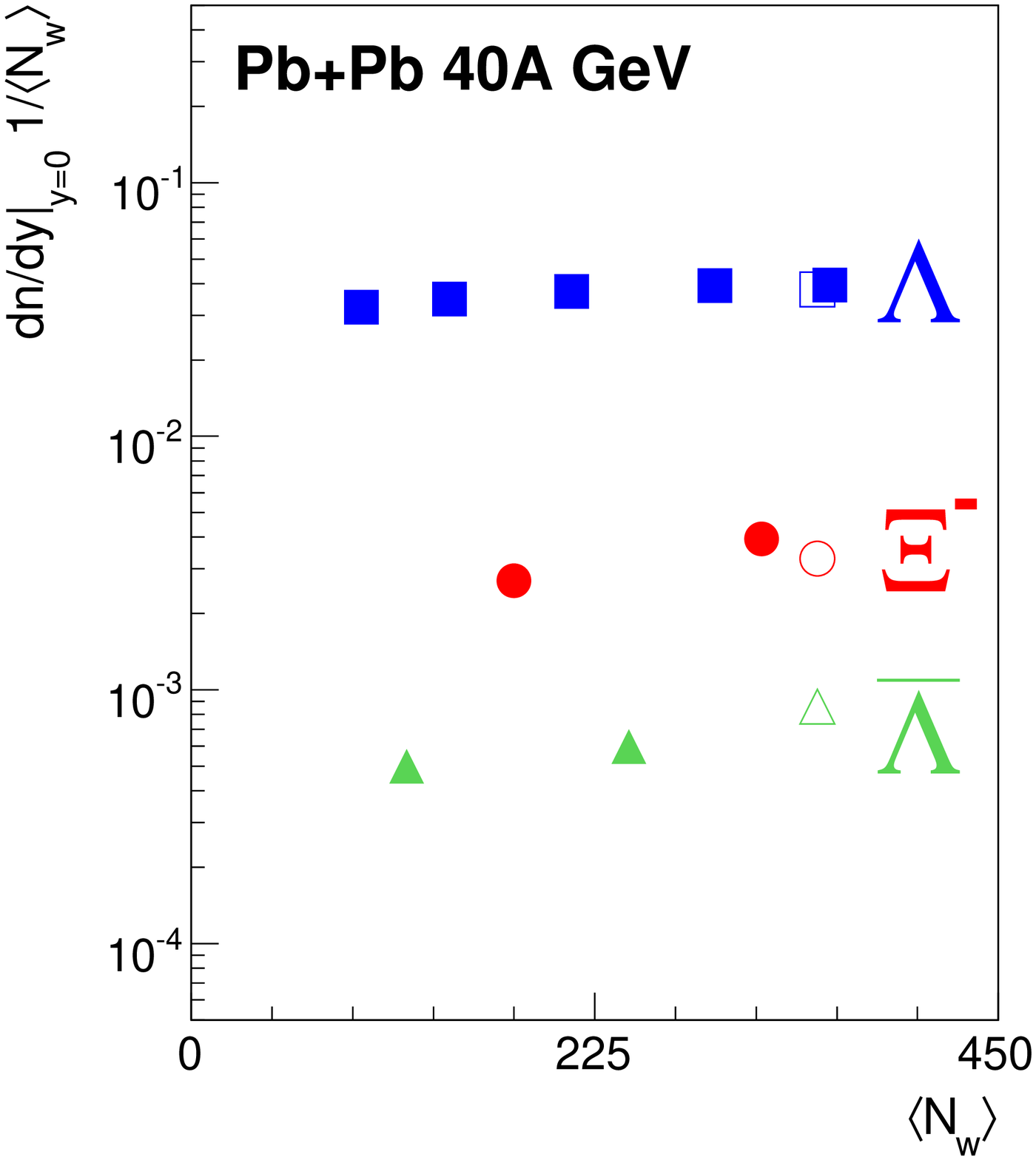}
\end{center}
\vspace{1mm}
\end{minipage}
\begin{minipage}[b]{50mm}
\begin{center}
\includegraphics[height=60mm]{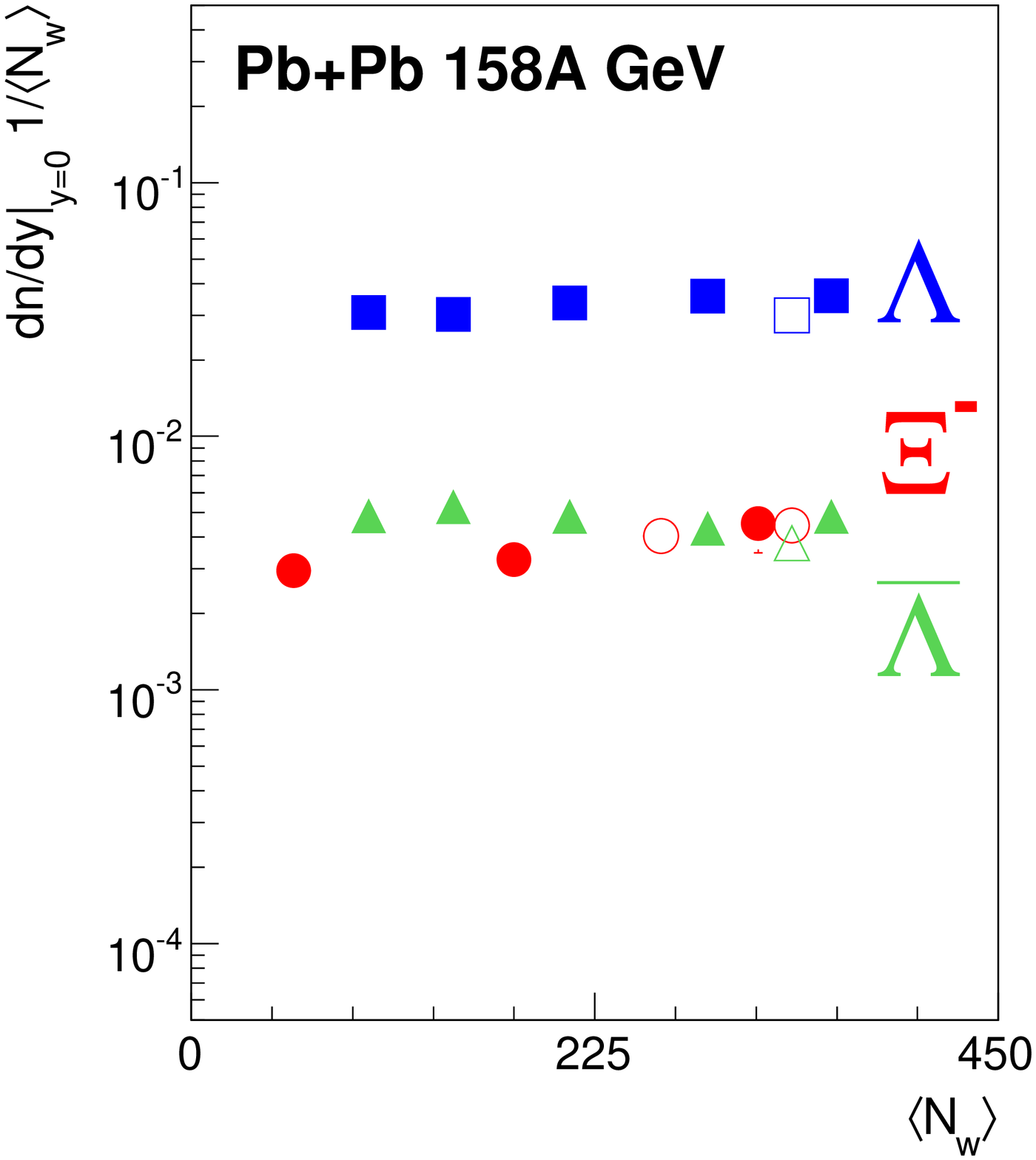}
\end{center}
\vspace{1mm}
\end{minipage}
\begin{minipage}[b]{50mm}
\begin{center}
\includegraphics[height=65mm]{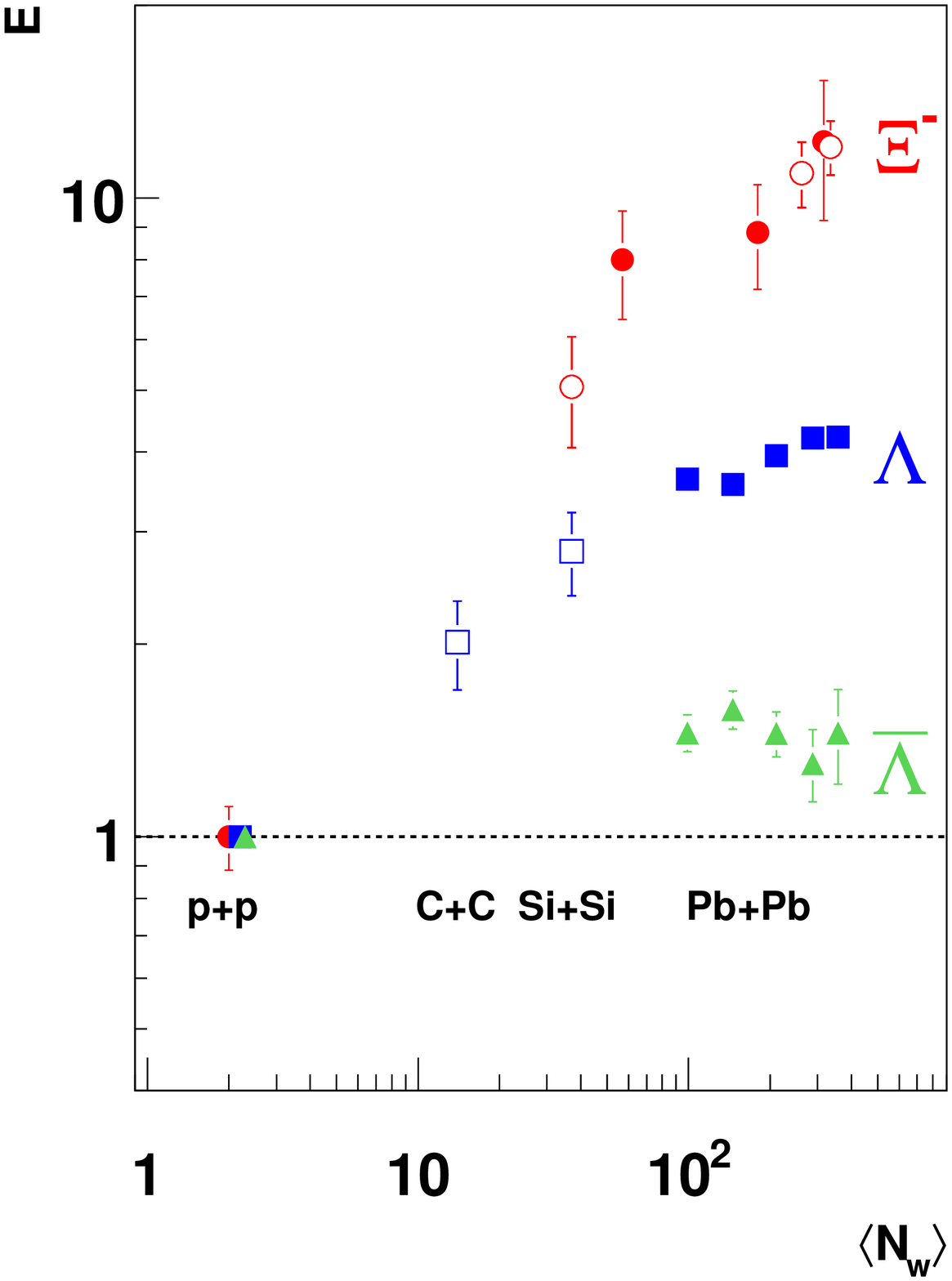}
\end{center}
\end{minipage}
\caption{The rapidity densities per wounded nucleon of \lam, \lab, 
and \xim\ at midrapidity ($|y| < 0.4$ for \lam\ (\lab), $|y| < 0.5$ 
for \xim) for minimum bias Pb+Pb collisions (filled symbols) as a 
function of \nwound.  Open symbols represent the results for online
selected central reactions.  
Right panel: The midrapidity yields per 
wounded nucleon relative to p+p yields for central C+C, Si+Si and minimum 
bias Pb+Pb reactions at 158\agev.}
\label{fig:hyperons}
\end{figure}

Figure~\ref{fig:hyperons} shows new NA49 results on \lam\ and \lab\ 
production in minimum bias Pb+Pb reactions at 40$A$ and 158\agev, 
together with preliminary data on \xim\ \cite{MICHI}.  The \lam\ data are 
corrected for feed down from weak decays.  While there is no 
system size dependence of the rapidity densities per wounded nucleon 
for \lam\ and \lab, a weak rise can be observed in the case of the \xim.  
In combination with measurements of yields in p+p, central C+C and Si+Si 
\cite{NA49SIZE} the system size dependence of the enhancement factor $E$, 
defined as:
\begin{equation}
E = \left. \left( \frac{1}{\langle N_{\rb{w}} \rangle} 
           \left. \frac{\textrm{d}N(\textrm{Pb+Pb})}{\textrm{d}y}\right|_{y=0}
    \right) \right/
    \left( \frac{1}{2} 
           \left. \frac{\textrm{d}N(\textrm{p+p})}  {\textrm{d}y}\right|_{y=0}
    \right)
\end{equation}
can be studied (rightmost panel of \Fi{fig:hyperons}).  
\nwound\ has been determined using the Glauber model \cite{NWOUND}.  The
enhancement exhibits a clear hierarchy ($E(\xim) > E(\lam) > E(\lab)$) 
and is almost independent on system size for $\nwound > 40$ for \lam\ 
and \lab.  For \xim\ a moderate \nwound\ dependence is seen.  Similar trends 
have been observed by the NA57 collaboration, however relative to a p+Be 
baseline \cite{NA57}.  Since already in p+A reactions a slight enhancement 
of strange particle production is observed \cite{NA49PA}, the enhancement 
relative to p+Be is less.

\section{Conclusions}

New data on (anti-)proton production at 20$A$~--~80\agev provide
insight into the energy dependence of stopping in the region where 
the onset of deconfinement possibly occurs.  No significant energy
dependence of \rdely\ is observed.  For the first time total
multiplicities of $^{3}$He have been measured, being in remarkable
agreement with statistical model predictions.  New results on the 
system size dependence of hyperon production allows to determine
the evolution of strangeness enhancement relative to elementary p+p
collisions.

\subsection*{Acknowledgments}

\begin{scriptsize}
This work was supported by the US Department of Energy
Grant DE-FG03-97ER41020/A000,
the Bundesministerium fur Bildung und Forschung, Germany, 
the Virtual Institute VI-146 of Helmholtz Gemeinschaft, Germany,
the Polish State Committee for Scientific Research (1 P03B 006 30, 
1 P03B 097 29, 1 PO3B 121 29, 1 P03B 127 30),
the Hungarian Scientific Research Foundation (T032648, T032293, T043514),
the Hungarian National Science Foundation, OTKA, (F034707),
the Polish-German Foundation, the Korea Science \& Engineering Foundation 
(R01-2005-000-10334-0) and the Bulgarian National Science Fund (Ph-09/05).
\end{scriptsize}



\end{document}